\documentclass[journal=jctcce,manuscript=article]{achemso}
\setkeys{acs}{maxauthors=0,articletitle=true}
\usepackage{achemso}
\setkeys{acs}{maxauthors=10, etalmode=truncate}
\usepackage{amssymb,amsfonts}
\usepackage{graphicx}
\usepackage{color}
\usepackage[table,dvipsnames]{xcolor}
\usepackage{amsmath}
\usepackage{bm}
\usepackage{braket}
\usepackage[version=4]{mhchem}
\usepackage{hyperref}
\usepackage{cleveref}
\usepackage{doi}

\definecolor{green}{HTML}{66c2a5}
\definecolor{orange}{HTML}{fc8d62}

\newcommand{\otr}{\texttt{OpenTrustRegion}}

\newcommand*\Eh{\ensuremath{\textrm{E}_\textrm{h}}}

\author{Jonas Greiner}
\email{jongr@dtu.dk}
\affiliation[dtu]{DTU Chemistry, Technical University of Denmark\\Kemitorvet Bldg. 206, 2800 Kgs. Lyngby, Denmark}
\author{Ida-Marie H{\o}yvik}
\email{ida-marie.hoyvik@ntnu.no}
\affiliation[ntnu]{Department of Chemistry, Norwegian University of Science and Technology\\Realfagbygget, D3-139, Trondheim 7491, Norway}
\author{Susi Lehtola}
\email{susi.lehtola@helsinki.fi}
\affiliation[helsinki]{Department of Chemistry, University of Helsinki\\P. O. Box 55, FIN-00014, Finland}
\author{Janus J. Eriksen}
\email{janus@dtu.dk}
\affiliation[dtu]{DTU Chemistry, Technical University of Denmark\\Kemitorvet Bldg. 206, 2800 Kgs. Lyngby, Denmark}

\title{A Reusable Library for Second-Order Orbital Optimization Using the Trust Region Method}

\begin{document}

\begin{abstract}

We present a reusable, open-source software implementation of the second-order trust region algorithm in the new \otr{} library.
We apply the implementation to the general-purpose optimization of molecular orbitals in various contexts within electronic-structure theory.
Our permissibly licensed implementation can be included in any software package, be it free and open-source, academically licensed closed-source, or commercial.
Detailing the implementation in \otr{}, we present a review of the theory behind trust region-based methods alongside various extensions.
We demonstrate the robustness and efficiency of our optimization library with extensive benchmarks for self-consistent field calculations, orbital localization, as well as orbital symmetrization tasks, featuring challenging and pathological systems.

\end{abstract}

\newpage

\begin{figure}[ht]
\begin{center}
\includegraphics[width=0.75\textwidth]{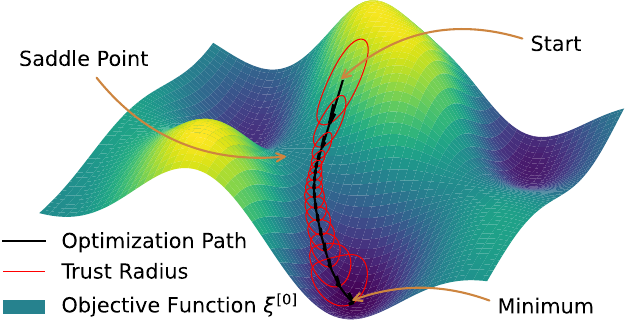}
\caption*{TOC graphic.}
\label{toc_fig}
\end{center}
\end{figure}

\newpage

\section{Introduction}\label{sec:intro}

Mean-field methods, such as Hartree--Fock (HF) and Kohn--Sham density functional theory (KS-DFT), serve as the fundamental models that underpin contemporary wave function-based electronic-structure theory. 
The optimization of the HF or KS-DFT energy onto an extremum---which should typically coincide with the global minimum of said functional---results in a self-consistent field (SCF) theory for the molecular orbitals (MOs)~\cite{mest, Parr1995}.

In order to be solved on a computer, the MOs, $\psi_{i}$, need to be discretized, which is most commonly done by expanding them as a linear combination of a set of basis functions, $\chi_{\mu}$,
\begin{equation}
  \label{eq:mo-expansion}
  \psi_i({\bm r}) = \sum_\mu C_{\mu i} \chi_\mu ({\bm r}) \ ,
\end{equation}
where ${\bm C}$ denote the MO expansion coefficients.
Most quantum-chemical applications employ Gaussian-type atomic orbitals (AOs), but the basis functions can also be chosen in other ways~\cite{Baerends1973_CP_41, Averill1973_JCP_6412, Payne1992_RMP_1045, Beck2000_RMP_1041, Genovese2008_JCP_14109, Blum2009_CPC_2175, White2017_JCP_244102, Lehtola2019_IJQC_25968}.
The problem of locating an extremum of a mean-field energy thus reduces to finding a set of expansion coefficients that satisfies a certain optimization condition, which, in turn, corresponds to solving a so-called Roothaan equation~\cite{Roothaan1951_RMP_69, hall_scf_prslsa_1951, Pople:1954aa, Berthier1954_JCP_363}, e.g., the spin-restricted variant:
\begin{equation}
 \label{eq:roothaan}
 \bm{F}(\bm{C})\bm{C} = \bm{S}\bm{C}\bm{E} \ .
\end{equation}
In Eq. \ref{eq:roothaan}, $\bm{F}$ denotes an AO Fock matrix that depends on the expansion coefficients, $\bm{C}$, while $\bm{S}$ is the overlap matrix between the AOs. 
At convergence, the orbitals will form a potential, ${\bm F}({\bm C})$, in which the MOs themselves are the lowest-lying eigenvectors. Starting from some initial guess, Eq. \ref{eq:roothaan} is then solved in an iterative manner until convergence is met~\cite{Wolfsberg1952_JCP_837, Hoffmann1963_JCP_1397, Jansik2009_PCCP_5805, Jansik2009_JCTC_1027, Norman2012_CPL_229, Zhang2016_IJQC_357, Lim2016_IJQC_1397, Lehtola2019_JCTC_1593}.

The present work is focused on an alternative approach.
Rather than solving a set of SCF equations, derived from the extremality condition of the corresponding energy functional, MOs can also be optimized by more direct means: starting from a set of guess orbitals, ${\bm C}_0$, one conducts a search over unitary transformations parameterized by an antisymmetric matrix, $\bm{\kappa}$, that minimizes the energy,
\begin{equation}
 \label{eqn:orbrot}
 E(\bm{\kappa}) = E(\bm{C}_0 e^{\bm{\kappa}}) \ .
\end{equation}
In Eq. \ref{eqn:orbrot}, the energy, $E(\bm{\kappa})$, is expressed as a function defined on the manifold of orthonormal orbitals where each point is reached by a unitary rotation of the reference orbitals. The task of solving the SCF equations of Eq. \ref{eq:roothaan} is now replaced by an optimization problem,
\begin{equation}
\label{eqn:optprob}
\bm{\kappa}^* = \arg \min E(\bm{\kappa}) \ ,
\end{equation}
which can be solved through iterative rotations of the given set of MOs. 
These orbitals do not lead to a fully diagonal Fock matrix like in Eq. \ref{eq:roothaan}, but they still ensure that the occupied-virtual blocks are zero and, thus, that the extremality condition of the total energy is satisfied.
These MOs can then be post-processed to recover the traditional set of canonical orbitals by diagonalizing the occupied-occupied and virtual-virtual blocks of the Fock matrix.

Beyond this type of optimization of mean-field energy functionals, which formally takes place on so-called Grassmannian manifolds, the basic rotation technique is useful also for other types of orbital optimizations.
The classical example of challenging orbital optimization is that of multiconfigurational self-consistent field theory~\cite{Levy1970_IJQC_297, Grein1971_CPL_44, Banerjee1976_IJQC_123, Ruedenberg1979_IJQC_1069, Dalgaard1978, Dalgaard1979, Lengsfield1980_JCP_382, Jorgensen1980, Shepard1980, Werner1980_JCP_2342, Yeager1980, Yeager1980a, Jorgensen1981, Lengsfield1981_JCP_478, Werner1981_JCP_5794, Igawa1982, Olsen1982_JCP_356, Olsen1982_JCP_6109, Yeager1982_JPC_2140, Lengsfield1982_JCP_4073, Jensen1984_CPL_140, Knowles1985_CPL_259, Werner1985_JCP_5053, Chaban1997_TCA_88, Angeli2002, Kreplin2019_JCP_194106, Kreplin2020_JCP_74102, Nottoli2021_JCTC_6819} (MCSCF), typically formulated in the complete active space (CAS) variant~\cite{Roos1980_CP_157, Roos1980_IJQC_175, Siegbahn1981_JCP_2384, Meier1989_TCA_95, Bates2015_JCP_44112, Hohenstein2015_JCP_224103, Lipparini2016_JCTC_4284, Sun2017_CPL_291, Reynolds2018_JCP_14106, Kollmar2019_JCC_1463}, where the optimization takes place on flag manifolds~\cite{Vidal2024_JPCA_6601}. The same is true for the geometric direct minimization of low-spin configuration state functions~\cite{burton_rohf_arxiv_2025}.
In addition, some quantum-chemical models require other types of strenuous optimization: some truncated coupled cluster models take place on so-called Stiefel manifolds~\cite{Hurley1953_PRSAMPES_446, Hunt1972_JCP_738, Ukrainskii1977_TMP_816, Goddard1978_ARPC_363, Cullen1996_CP_217, Beran2005_JPCA_92, Parkhill2009_JCP_84101, Parkhill2010_JCP_24103, Lehtola2018_MP_547, Lehtola2025_MP_2489096}, and so do rotations that intend to enforce either spatial locality~\cite{Lehtola2013_JCTC_5365} or point-group symmetries among orbitals by means of tailored unitary transformations~\cite{eriksen_symlo_jpca_2023}.

Direct optimization methods offer important conceptual advantages over SCF calculations.
For instance, these enable the control to avoid convergence onto saddle points, as we will demonstrate later in this work, and allow also for linear-scaling techniques that bypass the explicit MO diagonalization which otherwise hinders the exploitation of sparsity in the underlying AO basis~\cite{Hoest2008}.
Nonetheless, the procedure is far from simple in a numerical sense.

First, the optimization takes place in a very high-dimensional space.
The HF and KS-DFT energy functionals depend only on rotations between occupied and unoccupied (virtual) orbitals, which have $ov$ real angles for a set of $o$ occupied and $v$ virtual orbitals.
Nowadays, routine calculations typically employ thousands of basis functions, which implies that the corresponding optimization problems easily involve millions of degrees of freedom.
Second, the optimization manifold is profoundly corrugated: a $2 \times 2$ orbital rotation through an angle, $\theta$, returns an identity operation for $\theta=2k\pi$ ($k \in \mathbb N$).
Thus, while typical optimizations feature many simultaneous orbital rotations, the problem will still exhibit quasi-periodic character.
Third, any many-dimensional optimization is obviously hard: one tries to move down in energy as fast and securely as possible on a complicated function surface that is not globally known with only local information at one's disposal, e.g., slope and curvature. 

Pertinent to this last point is the fact that the search for extrema should try to avoid convergence onto first-order saddle points~\cite{HeadGordon1988_JPC_3063, cances_scf_conv_esaim_2000}; however, this is manifestly impossible if only information on the first derivative of the objective function at hand is known~\cite{Kudin2002_JCP_8255, Ziolkowski2008_JCP_204105, Seidl2022_JCTC_4164, Sethio2024_JPCA_2472, qin_smh_jctc_2024}.
For this reason, many so-called second-order methods that additionally employ second derivatives---either in exact or approximate form---have been proposed~\cite{Fischer1992_JPC_9768, Wong1995_JCC_1291, chaban_soscf_tca_1997, neese_soscf_cpl_2000, VanVoorhis2002_MP_1713, Sun2017__, Nottoli2021_MP_1974590, Dittmer2023_JCP_134104, Slattery2024_PCCP_6557}. 
However, in cases where the (exact) Hessian is not positive-definite, the Newton step will converge onto a saddle point instead of a local minimum even when such more elaborate and computationally expensive methods are used. 
As an example, false convergences and numerical issues can occur whenever the initial guess for the MOs happens to lie far away from the locally convex region.

As a remedy, a special class of methods emerges that instead operate in terms of only a limited region of trust~\cite{Fletcher:2000aa, More:1978aa, Sorensen1982_SJNA_409}, within which the objective function can be reliably approximated by a quadratic function~\cite{Bacskay1981_CP_385, Bacskay1982_CP_383, Simons:1983aa, Thogersen2004_JCP_16}.
The radius of this trust region can be dynamically updated during the optimization, and the restriction on the effective step length can prevent unwanted convergence onto saddle points~\cite{Joergensen1983_JCP_347, Jensen1984_JCP_1204, Jensen1986_CP_229, Jensen1987_JCP_451, Salek2007_JCP_114110, Host2008_JCP_124106, Hoyvik2012_JCTC_3137}.
Critically, these methods thus exhibit improved stability and robustness over a basic second-order Newton--Raphson algorithm.

To date, only a handful of software implementations of generally applicable trust region-based second-order algorithms have been published for general energy and orbital optimizations~\cite{Hoyvik2012_JCTC_3137, Folkestad:2022aa, Matveeva:2023aa, HelmichParis2021_JCP_164104, HelmichParis2022_JCP_204104, Feldmann2023_JCTC_856, Sepehri2023_JPCA_5231}. 
In our opinion, there are two likely main explanations for this scarcity.

First, not enough attention is paid to the possible shortcomings of traditional first- and second-order methods; for instance, the aforementioned risk of convergence onto saddle point solutions can only be detected if a rigorous stability check is performed {\textit{a posteriori}}, yet such analyses are typically not run by default in most electronic-structure program packages.

Second, the implementation of trust region-based methods admittedly demands considerable efforts, and the values of the associated hyperparameters are often incompletely documented, making verification and tuning by other parties a significant and real challenge.
This is a core issue of the traditional model behind scientific (academic) software development that has, to a large extent, acted as a motivation for the present work.
Historically, key features of our community codes have been engineered from scratch time and time again in processes that are arguably both repetitive, laborious, and often prone to errors~\cite{Lehtola2023_JCP_180901}.

To that end, modern version control systems have helped facilitate the development and distribution of source codes under decentralized global settings~\cite{Lehtola2022_WIRCMS_1610}.
As an example of an arguably more open approach, the recent \texttt{OpenOrbitalOptimizer} project features reusable implementations of the traditional SCF methodology~\cite{Lehtola2025_JPCA_5651, Lehtola2025__}, driven by the realization that the solution of Eq. \ref{eq:roothaan} only requires manipulations of matrices that are agnostic to the underlying conventions for the basis functions or type of basis set when operating in a MO basis.

In the same vein, the present work reports on our design of \otr{}~\cite{opentrustregion}, a reusable and open-source implementation of the second-order trust region algorithm applicable for general orbital optimizations.
Alike \texttt{OpenOrbitalOptimizer}, the optimization problem in \otr{} can be formulated in any orbital basis since the actual functions that define the optimization surface are implemented within the scope of a host program.
As a result, the same base implementation of the algorithm can be used across different program packages, regardless of how the MOs are discretized.
However, as \otr{} is a general-use optimizer, it may also be used to optimize non-orthonormal orbital expansions.

The \otr{} library is designed with an intention to be more versatile, more robust, and more reusable than any previously reported implementation of the second-order trust region method aimed for electronic-structure calculations.
To that end, the present work reports on interfaces to three different host programs, all written in different programming languages, with applications to the optimization of orbitals in the context of both HF and KS-DFT alongside spatial localization and point-group symmetrization of orbitals, thus allowing for a coherent and consistent workflow around these individual steps.

The layout of our work is as follows.
In Sect. \ref{sec:theory}, we revisit the general theory behind the trust region method.
We present key algorithmic details in Sect. \ref{sec:algorithm} and discuss the design of our software implementation in \otr{} in Sect. \ref{sec:implementation}.
In Sect. \ref{sec:results}, we present results for restricted and unrestricted SCF calculations on a benchmark set of pathological systems, comparing the performance to alternative state-of-the-art second-order solvers implemented in other program packages. 
We furthermore employ our algorithm in the localization and symmetrization of occupied and virtual MOs of extended systems in routine basis sets using a variety of cost functions.
Finally, we provide a summary and some conclusive remarks in Sect. \ref{sec:summary}, which also discusses additional potential applications of the presented library.

\section{Theory}\label{sec:theory}

In this section, we review the fundamental theory behind trust region-based orbital optimization. Particular emphasis is placed on modifications that allow for the algorithm to be agnostic to the type of objective function involved. 
We limit our discussion to minimization, as the maximization of $f(\bm{x})$ can be trivially reformulated as the minimization of $-f({\bm x})$.

\subsection{Quadratic Model}\label{sec:quad_model}

In the following, we will consider a general scalar objective function, $\xi$, that depends on some parameters, $\bm{\kappa}$, which correspond to orbital rotation angles throughout the present work.
That being said, $\bm{\kappa}$ could equally well be nuclear coordinates and the presented theory and discussion would remain almost exactly the same.
However, since the number of orbital rotation angles can easily surpass one million, the focus on orbital optimization in this work motivates many of the basic algorithmic design choices behind \otr{}.

Now, consider a Taylor series expansion of $\xi(\bm{\kappa})$ centered around $\bm{\kappa} = \bm{0}$,
\begin{equation}
\label{eq:xi-expansion}
\xi(\bm{\kappa}) = \xi^{[0]}+\bm{\kappa}^T\bm{\xi}^{[1]}+\frac{1}{2}\bm{\kappa}^T\bm{\xi}^{[2]}\bm{\kappa}+ \cdots \ ,
\end{equation}
where $\xi^{[0]}$ is the function value evaluated at $\bm{\kappa}=\bm{0}$, while $\bm{\xi}^{[1]}$ and $\bm{\xi}^{[2]}$ are the gradient and Hessian of $\xi(\bm{\kappa})$ with respect to $\bm{\kappa}$, respectively, both likewise evaluated at $\bm{\kappa} = \bm{0}$. 

Note that studying an expansion at $\bm{\kappa}=\bm{0}$ is not a limitation, given how the optimization proceeds iteratively on the manifold of orthonormal orbitals. 
At each iteration, the displacement parameters, $\bm{\kappa}$, define a tangent-space direction at the current point, and the corresponding update is realized through an orbital rotation via the exponential map in Eq. \ref{eqn:orbrot}.
This way, the accumulated effect of successive rotations corresponds to following a trajectory on the orbital manifold while preserving orthonormality throughout the optimization. 
The accumulation of $\bm{\kappa}$ is accomplished in the scope of our library by rotating the reference orbitals through the angle at every involved step in the course of the optimization procedure. 
Importantly, this approach is not limited to orthonormal orbital manifolds, e.g., for optimizations in 
$\mathbb{R}^n$, the accumulated displacement parameters can be added directly to follow the optimization trajectory.

By truncating the expansion of the objective function in Eq. \ref{eq:xi-expansion} to second order in $\bm{\kappa}$, one obtains the standard quadratic model which reads as 
\begin{equation}
\label{eq:quadratic_model}
\Lambda(\bm{\kappa})=\xi^{[0]}+\bm{\kappa}^T\bm{\xi}^{[1]}+\frac{1}{2}\bm{\kappa}^T\bm{\xi}^{[2]}\bm{\kappa} \ .
\end{equation}
Stationary points are found by setting $\nabla_{\bm{\kappa}} \Lambda(\bm{\kappa}) = \bm{0}$, yielding the Newton--Raphson equation:
\begin{equation}
\label{eq:nr_eq}
    \bm{\xi}^{[2]}\bm{\kappa}=-\bm{\xi}^{[1]} \ .
\end{equation}
The optimal step is now formally obtained from Eq. \ref{eq:nr_eq} as $\bm{\kappa} = -(\bm{\xi}^{[2]})^{-1}\bm{\xi}^{[1]}$. However, for Eq. \ref{eq:quadratic_model} to converge to a minimum, the Hessian must remain positive definite throughout the optimization. 
Yet, in most applications, this is initially not true, in which case the quadratic model will not have a minimizer, and one may risk optimizing onto a saddle point instead.

\subsection{Trust Region}
\label{sec:trust_region_full_space}

To circumvent the problem of possible negative Hessian eigenvalues, one may choose to constrain the step size, $h$, that is, the norm of $\bm{\kappa}$ in each optimization step, to a so-called region of trust.
As discussed in Ref. \citenum{Sorensen1982_SJNA_409}, this restriction on the step size is formally equivalent to employing a positive semi-definite, level-shifted Hessian in the Newton--Raphson equation of Eq. \ref{eq:nr_eq}.
Beyond addressing potential problems associated with negative Hessian eigenvalues, the proposed constraint brings about another added benefit---namely, it can be used to ensure that the quadratic model stays a good approximation to $\xi(\bm{\kappa})$ with coinciding minima.

The restriction on the step size, $h$, is conveniently introduced via a Lagrangian reformulation of Eq. \ref{eq:quadratic_model}, in which the multipliers, $\mu$, denote a so-called level shift:
\begin{equation}
\label{eq:lagrangian}
    L(\bm{\kappa}, \mu) =\Lambda(\bm{\kappa})-\frac{1}{2}\mu(\bm{\kappa}^T\bm{\kappa}-h^2) \ . 
\end{equation}
Differentiation of Eq. \ref{eq:lagrangian} with respect to $\mu$ now yields the step size (the $L^2$-norm of $\bm{\kappa}$),
\begin{equation}
\label{eq:stepsize_constraint}
    \bm{\kappa}(\mu)^T\bm{\kappa}(\mu)=\lVert \bm{\kappa}(\mu) \rVert^2 =h^2 \ ,
\end{equation}
while differentiation with respect to $\bm{\kappa}$ yields a level-shifted Newton--Raphson equation,
\begin{equation}
\label{eq:level_shifted_newton}
    (\bm{\xi}^{[2]}-\mu\bm{1})\bm{\kappa}(\mu)=-\bm{\xi}^{[1]} \ .
\end{equation}
Eq. \ref{eq:level_shifted_newton} thus implies that the optimal rotation angle will depend explicitly on the level shift,
\begin{equation}
 \label{eq:kappa-mu}
    \bm{\kappa}(\mu)=-(\bm{\xi}^{[2]}-\mu\bm{1})^{-1}\bm{\xi}^{[1]}\ .
\end{equation}
If $\mu$ takes a (negative) value less than the smallest Hessian eigenvalue, $\varepsilon_1$, the level-shifted Hessian in Eq. \ref{eq:kappa-mu} remains positive definite and the step leads towards a minimum of the constrained model. However, $\mu$ should not be chosen to be too negative, as Eq. \ref{eq:kappa-mu} then approaches a steepest descent step, which will result in slow convergence,
\begin{align}
\lim\limits_{\mu\to -\infty} \Vert\bm{\kappa}(\mu)\Vert &= 0 \ ,
\end{align}
and nor should $\mu$ lie too close to $\varepsilon_1$ since the step can then end up arbitrarily large, 
\begin{align}
\lim\limits_{\mu\to \varepsilon_1} \Vert\bm{\kappa}(\mu)\Vert &= \infty \ .
\end{align}
In fact, $\Vert\bm{\kappa}(\mu)\Vert$ exhibits a characteristic pole structure whenever $\mu$ matches any eigenvalue (cf. Fig. 1 of Ref. \citenum{Hoyvik2012_JCTC_3137}). 
The width and shape of these poles depend on the off-diagonal blocks of the Hessian, so knowledge of only the eigenvalues of this matrix is generally insufficient for choosing a safe and effective value of $\mu$. 
However, as $\Vert\bm{\kappa}(\mu)\Vert$ increases monotonically in the interval below the first pole, the level shift that corresponds to a prescribed trust region radius can be determined via root-finding algorithms, such as, the bisection method.

\subsection{Augmented Hessian}

Rewriting Eq. \ref{eq:lagrangian} in matrix form yields the following expression for the Lagrangian:
\begin{equation}
\label{eq:augmented_hessian_lagrangian}
    L(\bm{\kappa}, \mu)=\xi^{[0]}+\frac{1}{2}\begin{pmatrix}
       1 \\
       \bm{\kappa}
    \end{pmatrix}^T\begin{pmatrix}
       0 & {\bm{\xi}^{[1]}}^T \\
       \bm{\xi}^{[1]} & \bm{\xi}^{[2]}
    \end{pmatrix}\begin{pmatrix}
       1 \\
       \bm{\kappa}
    \end{pmatrix}-\frac{1}{2}\mu\left(\begin{pmatrix}
       1 \\
       \bm{\kappa}
    \end{pmatrix}^T\begin{pmatrix}
       1 \\
       \bm{\kappa}
    \end{pmatrix}-h^2-1\right) \ .
\end{equation}
Differentiation with respect to $\bm{\kappa}$ yields an eigenvalue equation for an augmented Hessian,
\begin{equation}
    \label{eq:unscaled_augmented_hessian_eigenvalue}
    \begin{pmatrix}
       0 & {\bm{\xi}^{[1]}}^T \\
       \bm{\xi}^{[1]} & \bm{\xi}^{[2]}
    \end{pmatrix}\begin{pmatrix}
           1 \\
           \bm{\kappa}
         \end{pmatrix}
        =\mu\begin{pmatrix}
           1 \\
           \bm{\kappa}
         \end{pmatrix} \ ,
\end{equation}
while differentiation with respect to $\mu$ again yields Eq. \ref{eq:stepsize_constraint}.
These conditions must be solved simultaneously, also given that $\mu$ determines $\bm{\kappa}$ via Eq. \ref{eq:kappa-mu}.

However, an elegant solution has been suggested by Jensen and J{\o}rgensen~\cite{Jensen1984_JCP_1204}, in which the gradient is multiplied by a scalar, $\alpha > 0$. Denoting $\bm{x}(\alpha)=\alpha \bm{\kappa}(\mu)$, Eq. \ref{eq:unscaled_augmented_hessian_eigenvalue} now reads as
\begin{equation}
\label{eq:augmented_hessian_eigenvalue}
    \begin{pmatrix}
       0 & \alpha{\bm{\xi}^{[1]}}^T \\
       \alpha\bm{\xi}^{[1]} & \bm{\xi}^{[2]}
    \end{pmatrix}\begin{pmatrix}
           1 \\
           \bm{x}(\alpha)
         \end{pmatrix}
        =\mu\begin{pmatrix}
           1 \\
           \bm{x}(\alpha)
         \end{pmatrix} \ ,
\end{equation}
allowing for $\bm{\kappa}$ and $\mu$ to be solved by finding the scaling factor, $\alpha$, that satisfies Eq. \ref{eq:stepsize_constraint}.

Now, as per the Hylleraas--Undheim--MacDonald theorem~\cite{hylleraas_1930, mac_donald_1933}, the smallest eigenvalue of the scaled augmented Hessian in Eq. \ref{eq:augmented_hessian_eigenvalue} will be a lower bound for the smallest eigenvalue of the true Hessian, whatever the value of $\alpha$.
The augmented Hessian procedure thus automatically ensures a level shift, $\mu$, that leads to the desired positive-definite, level-shifted Hessian, while also yielding $\bm{\kappa}$ of the correct length.
If the Hessian itself is positive definite and the step is inside the trust region, a level shift is not needed as the Newton step can be trusted inside the convex region of the optimization surface, thus leading to a minimum.

\section{Algorithm}\label{sec:algorithm}

In the following section, we present our algorithm for solving the general trust region method outlined in Sect. \ref{sec:theory}.
In-depth details are provided on key technical subtleties, such as, the involved Davidson procedures, the choice of trial vectors, the employed preconditioning techniques, and the algorithms used to update both the level shift and trust region. 
Readers with a greater interest in aspects related to our reusable implementation are encouraged to skip ahead directly to Sect. \ref{sec:implementation}.
Table S1 of the supporting information (SI) presents a full list of all the hyperparameters used in \otr{} alongside their numerical values.

\subsection{Davidson Algorithm}\label{sec:davidson_algo}

Each macroiteration of the overall procedure involves the diagonalization of the augmented Hessian matrix in Eq. \ref{eq:augmented_hessian_eigenvalue}.
Given that this matrix is built from the full Hessian, $\bm{\xi}^{[2]}$, the row and column dimensions of which scale quadratically with the number of MOs, a full diagonalization is obviously intractable in the general case.
For this reason, we employ a Davidson algorithm to determine the lowest-lying eigenvectors of Eq. \ref{eq:augmented_hessian_eigenvalue} along with the optimal level shift, $\alpha$, in a reduced space~\cite{Davidson1975_JCP_87}.
The iterative algorithm works in terms of a set of trial vectors, $\{\bm{b}_i\}$, the first of which is chosen as the normalized gradient,
\begin{equation}
\label{eq:first_trial}
    \bm{b}_1=\frac{\bm{\xi}^{[1]}}{\lVert\bm{\xi}^{[1]}\rVert} \ .
\end{equation}

With this choice, the augmented Hessian eigenvalue equation in the reduced space stays on par with the general form in Eq. \ref{eq:augmented_hessian_eigenvalue}; the trial vectors may then be separated into a direction parallel to the gradient ($\bm{b}_1$) and along its orthogonal complements, {\textit{vide infra}}.

In the reduced space of the trial vectors, the augmented Hessian is then expressed as:
\begin{equation}
    \bm{A}(\alpha)
    =\begin{pmatrix}
       0 & \alpha\lVert\bm{\xi}^{[1]}\rVert & 0 & 0 & \cdots \\
       \alpha\lVert\bm{\xi}^{[1]}\rVert & \bm{b}_1^T\bm{\sigma}_1 & \bm{b}_1^T\bm{\sigma}_2 & \bm{b}_1^T\bm{\sigma}_3 & \cdots \\
       0 & \bm{b}_2^T\bm{\sigma}_1 & \bm{b}_2^T\bm{\sigma}_2 & \bm{b}_2^T\bm{\sigma}_3 & \cdots \\
       0 & \bm{b}_3^T\bm{\sigma}_1 & \bm{b}_3^T\bm{\sigma}_2 & \bm{b}_3^T\bm{\sigma}_3 & \cdots \\
       \vdots & \vdots & \vdots & \vdots & \ddots \\
    \end{pmatrix} \ . \label{eq:a_matrix}
\end{equation}
Here, $\bm{\sigma}$ is used to denote the general action of the Hessian on the basis vectors,
\begin{equation}
\label{eq:linear_transformation}
\bm{\sigma}_i = \bm{\xi}^{[2]}\bm{b}_i \ ,
\end{equation}
which is the core operation required for the second-order solution.
Eq. \ref{eq:augmented_hessian_eigenvalue} now reads as
\begin{equation}
\label{eq:eigenvalue_problem_reduced}
    \bm{A}(\alpha)\begin{pmatrix}
           1 \\
           \bm{x}^\mathrm{red}(\alpha)
         \end{pmatrix}
        =\mu\begin{pmatrix}
           1 \\
           \bm{x}^\mathrm{red}(\alpha)
         \end{pmatrix} \ ,
\end{equation}
where $\bm{x}^\mathrm{red}$ denote the expansion coefficients of ${\bm x}$ in the linear combination of reduced-space basis functions.
The dimension of this problem is generally small, as determined by the number of trial vectors.
The reduced-space matrices are easy to construct and diagonalize, and one needs only to determine the action of the Hessian on the trial vectors, cf. Eq. \ref{eq:linear_transformation}.

The solution to Eq. \ref{eq:eigenvalue_problem_reduced} in the reduced space now provides an approximation to the level-shifted Newton equation in the full space,
\begin{equation}
\label{eq:full_space_step}
    \bm{\kappa}(\mu)=\frac{1}{\alpha}\sum_{i=1}^nx_i^\mathrm{red}(\alpha)\bm{b}_i \ , 
\end{equation}
and the residual of the solution is computed according to the following expression:
\begin{equation}
\label{eq:residual}
\begin{split}
  \bm{R}_n &=  -\bm{\xi}^{[1]} -(\bm{\xi}^{[2]}-\mu\bm{1})\frac{1}{\alpha}\sum_{i=1}^nx_i^\mathrm{red}(\alpha)\bm{b}_i\\
 &= -\bm{\xi}^{[1]} -\frac{1}{\alpha}\sum_{i=1}^nx_i^\mathrm{red}(\alpha)\bm{\sigma}_i+ \frac{\mu}{\alpha}\sum_{i=1}^n x_i^\mathrm{red}(\alpha)\bm{b}_i \ .
\end{split}
\end{equation}
If the norm of this residual falls below a given threshold, the microiterations have converged and orbitals are updated from the parameters in Eq. \ref{eq:full_space_step};
if not, a new trial vector, $\bm{b}_{n+1}$, is generated.

The traditional and arguably most simple way to generate trial vectors is to orthogonalize the residual against all previous trial vectors via a standard Gram--Schmidt procedure.
However, preconditioning techniques are useful for accelerating convergence; for instance, the inverse of the level-shifted Hessian diagonal offers an effective and convenient choice:
\begin{equation}
\label{eq:diagonal_preconditioner}
\bm{P}=\left(\bm{\xi}^{[2]}_\mathrm{diag}-\mu\bm{1}\right)^{-1} \ .
\end{equation}
New trial vectors are thus generated by preconditioning of Eq. \ref{eq:residual},
\begin{equation}
    \label{eq:precond_residual}
    \tilde{\bm{b}}_{n+1}=\bm{P}\bm{R}_n - \sum_{i=1}^{n} (\bm{P} \bm{R}_{n} \cdot \bm{b}_{i}) \bm{b}_{i} \ ,
\end{equation}
followed by regular orthonormalization~\bibnote{We note that the diagonal preconditioner in Eq. \ref{eq:diagonal_preconditioner} is just one choice among many more elaborate options. 
However, we have found this choice to be generally effective for our purposes in all cases where the Hessian matrix is diagonally dominant. 
Furthermore, given that we also rely on the diagonal Hessian for generating start vectors (see Sect. \ref{sec:initial-vectors}), the preconditioner is readily formed in the algorithm without requiring access to any additional data.}. 
Next, the corresponding sigma, $\bm{\sigma}_{n+1}$, and residual, $\bm{R}_{n+1}$, vectors are determined and the procedure is repeated until convergence is met.

\subsection{Initial Trial Vectors \label{sec:initial-vectors}}

While a typical Davidson procedure is initiated from just the single trial vector of Eq. \ref{eq:first_trial}, our algorithm includes an option for augmenting the starting guess by additional basis vectors.
The possible inclusion of these is motivated by two distinct reasons.

First, in the case of orbital localization, the eigenvalue spectrum of the Hessian often exhibits a large spectral range, and the use of a single initial guess vector may result in the lowest roots not being found.
Fortunately, such excessively large negative values can typically also be found along the diagonal of the Hessian, and to ensure that the Davidson procedure indeed locates the lowest root also in such cases, we add a unit vector, ${\bm{e}} = (0, \dots, 1, \dots, 0)^\text{T}$, with the non-zero entry at the location of the smallest element of the Hessian diagonal~\cite{hoyvik_trust_region_jctc_2012}.

Second, given how our algorithm relies on Hessian-vector products, point-group symmetries can also cause numerical problems: the gradient between orbitals of different symmetries must vanish on account of symmetry, which may result in the Davidson procedure not being able to find the lowest roots of the Hessian.
In these cases, a symmetry-constrained solution will correspond to a saddle point rather than a minimum.
To circumvent this potential issue, particularly for molecular systems belonging to high and degenerate point groups, we provide the option to augment the initial set of trial vectors by $N_\textrm{random}$ random basis vectors.

\subsection{Level Shift Update}
\label{sec:dynamic_update_mu}

Knowledge about the full space is gradually built up in the Davidson iterations by adding new trial vectors in each microiteration. 
Thus, for the level shift to always stay below the lowest eigenvalue of the augmented Hessian in the reduced space, it must be redetermined at each microiteration by dynamically adapting to the information present in $\bm{A}$, cf. Eq. \ref{eq:a_matrix}. 
As discussed in Sect. \ref{sec:trust_region_full_space}, it is the role of the scaling factor, $\alpha$, to control the step length.
Since the norm of $\bm{\kappa}(\mu)$ can be shown to be monotonically increasing for $\mu < \epsilon_1$ (the lowest Hessian eigenvalue), we do a bisectional search in the $\alpha$ parameter to identify the point where
\begin{equation}
    \label{eq:norm_h}
    \Vert\bm{x}^\mathrm{red}(\alpha)\Vert\approx h \ .
\end{equation}

This search amounts to solving the reduced-space eigenvalue equation of Eq. \ref{eq:eigenvalue_problem_reduced} for several values of $\alpha$. 
The parameter that fulfills Eq. \ref{eq:norm_h} will yield the level shift of the current Davidson iteration as the corresponding eigenvalue of Eq. \ref{eq:eigenvalue_problem_reduced}.
In each microiteration, if the lowest eigenvalue of the reduced-space augmented Hessian is above a threshold (e.g., $-10^{-5}$), the algorithm simply performs a Newton–Raphson step and sets $\mu=0$. In this case, no bisection and diagonalization of the reduced Hessian is necessary. Microiterations proceed until the residual norm falls below the convergence threshold, as usual.
Even though the eigenvalue equation is thus solved multiple times during a microiteration, no computational overhead is incurred in practice due to the small dimension of the reduced subspace. 

\subsection{Trust Region Update}\label{sec:trust_region_update}

When the microiterations have converged to a proposed step, it describes the minimum on the given boundary of the trust region.
The step is thus accepted if it leads to a decrease in the value of the objective function and rejected otherwise.
The trust-region radius is then adjusted according to how well the actual reduction agrees with the quadratic model in Eq.~\ref{eq:quadratic_model}: if the step reduces the objective function significantly more than predicted, the radius is increased, whereas if it reduces it less or leads to an increase, the trust radius is decreased.
Specific thresholds and scaling factors for these updates are reported in Table S1 of the SI.

If the objective function is cheap to evaluate---which is typically the case in orbital localization---one may optionally perform a full line search along the $\bm{\kappa}$ vector produced by the trust region solver at this stage, potentially leading to a larger decrease in function value.

\section{Implementation}\label{sec:implementation}

The trust region procedure outlined in Sects. \ref{sec:theory} and \ref{sec:algorithm} has been implemented in a new library, \otr{}, written in Fortran for optimal compatibility across different host programs~\cite{opentrustregion}.
All of the essential functionality is contained within a single source file to be embedded into existing software projects on account of its permissive Mozilla Public License.

The \otr{} library further provides interfaces to the C and Python programming languages to ensure broad interoperability.
In addition to being used as an embedded Fortran library, the solver may thus also be accessed as a dynamic library from codes written in other languages and as a module directly from within Python code.
To demonstrate these features, we have successfully interfaced it with both the \texttt{PySCF}~\cite{sun_pyscf_2017, Sun2020_JCP_24109}, \texttt{Psi4}~\cite{Turney2012_WIRCMS_556, Parrish2017_JCTC_3185, Smith2020_JCP_184108}, and \texttt{eT}~\cite{Folkestad2020_JCP_184103} program packages that cover three of the most widely used programming languages in quantum chemistry (Python, C++, and Fortran, respectively).
These interfaces offer support for restricted closed- (R-) and open-shell (RO-) as well as unrestricted (U-) calculations at the level of both HF and KS-DFT, either with or without the use of point-group symmetry.
Furthermore, \otr{} can be used for generalized Pipek--Mezey~\cite{Pipek1989_IJQC_487, Pipek1989_JCP_4916,Lehtola2014_JCTC_642} (gPM), Foster--Boys~\cite{Foster1960_RMP_300} (FB), and Edmiston--Ruedenberg~\cite{Edmiston1963_RMP_457, Edmiston1965_JCP_97} (ER) orbital localization, depending only on the availability of appropriate orbital Hessian implementations in any specific choice of host program.
Finally, spatially localized MOs can be symmetrized with respect to general (non-)Abelian point groups via an interface to the recent {\texttt{SymLo}} code~\cite{symlo}.

\begin{figure}
    \centering
    \includegraphics[width=\linewidth]{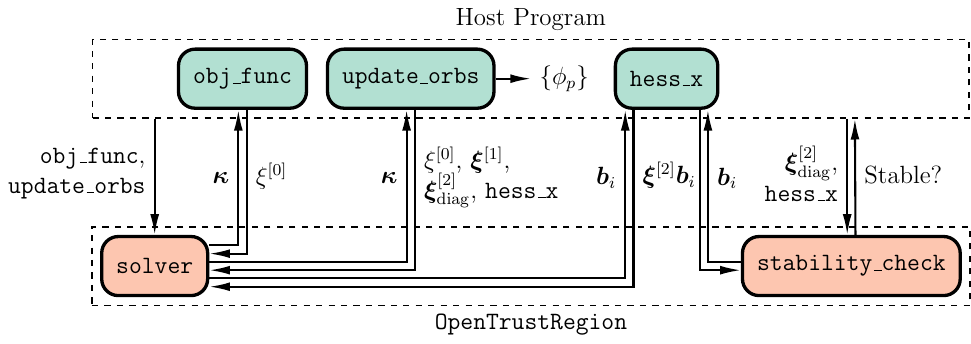}
    \caption{Interface between a given host program and the \otr{} library. \label{fig:interface}}
\end{figure}

Communication between a host program of choice and the \otr{} library is handled via function pointers. As per the generic interface design illustrated in Fig. \ref{fig:interface}, the following functions must be exposed and provided by the host:
\begin{enumerate}
  \item \texttt{obj\_func}: Evaluate the objective function for a given rotation, $\bm{\kappa}$: $\xi = \xi(\bm{\kappa})$
  \item \texttt{update\_orbs}: Transform orbitals through $\bm{\kappa}$: ${\bm C} \to {\bm C} e^{\bm{\kappa}}$. The function should collectively return the new function value, $\xi^{[0]}$, gradient, $\bm{\xi}^{[1]}$, and Hessian diagonal, $\bm{\xi}^{[2]}_{\textrm{diag}}$, alongside a custom function (\texttt{hess\_x}) for evaluating Hessian-vector products    
  \item \texttt{hess\_x}: Compute the sigma vectors of Eq. \ref{eq:linear_transformation}
\end{enumerate}
The supplied callables are responsible for applying any relevant projections to enforce constraints and for ensuring that only non-redundant parameters are passed to the library.
Examples of such constraints include (but are not restricted to) orthogonality, spin symmetry, and point-group symmetry.
This flexible design will thus allow for the reusable application of second-order trust region optimization techniques to a wide range of objective functions under very general constraints.
Additionally, but as an integral component, our library can be used to perform both internal and external stability checks of the obtained solution.
The stability check can also provide the negative curvature direction, which can be used to steer away from a saddle point.
Using an optional boolean argument, the internal stability check can be performed automatically whenever the solver reaches a stationary point, carrying out a line search along the negative curvature direction and restarting the optimization procedure, if necessary.
In practice, however, we have not observed our implementation to converge onto saddle points, meaning that the underlying algorithm robustly and consistently reaches true local minima.
External stability checks can be performed by calling the stability check with the modified Hessian of a less constrained wave function. 

As was discussed in Sect. \ref{sec:algorithm}, our algorithm is implemented in a manner that deliberately avoids the storage of the full Hessian, instead relying solely on repeated Hessian-vector products.
The \texttt{hess\_x} function thus takes a trial vector as input and contracts this with the rows of the Hessian on-the-fly to compute the Hessian-vector product, thereby avoiding the explicit storage of individual Hessian elements.
The diagonal preconditioner of Sect. \ref{sec:davidson_algo} is used by default within the Davidson procedure, although our library offers ample flexibility for the host program to supply alternative preconditioners as an optional callable.
For the optional line search procedure after each macroiteration (cf. Sect. \ref{sec:trust_region_update}), we use the standard numerical recipe outlined in Ref. \citenum{Press1992__}.

\section{Results \label{sec:results}}

\subsection{Computational Details}\label{sec:comp_details}

All results to follow were obtained employing our interface to {\texttt{PySCF}}.
Using the same initial guesses and tight convergence criteria, the results obtained using \otr{} via interfaces to other external programs, e.g., \texttt{Psi4} and \texttt{eT}, will (and are confirmed to) agree up to numerical differences arising only from the use of finite floating-point arithmetic.

For the SCF calculations in Sect. \ref{sec:scf_results}, results are presented on a small set of pathological systems: Ni$_2$, NiC, Cr$_2$, CrC, Ti$_2$O$_4$, UF$_4$, and UO$_2$(OH)$_4$. 
The geometries for the Cr$_2$, CrC, UF$_4$, and UO$_2$(OH)$_4$ systems are taken from Ref. \citenum{Daniels2000_PCCP_2173}, Ni$_2$ and NiC use the same bond lengths as Cr$_2$ and CrC, whereas the geometry for Ti$_2$O$_4$ is that used in Ref. \citenum{Nottoli2021_MP_1974590}. 

The cc-pVTZ basis set~\cite{Dunning1989_JCP_1007, Balabanov2005_JCP_64107} is employed for the transition metal compounds, while the LANL2DZ basis set~\cite{hay_lanl2dz_1983} is used for the systems containing uranium.
The choice of LANL2DZ for UF$_4$ and UO$_2$(OH)$_4$ is due only to its earlier use in existing literature on SCF convergence problems, and is not to be interpreted as support for continued use of this basis set.

The PBE~\cite{Perdew1996_PRL_3865, Perdew1997_PRL_1396}, B3LYP~\cite{Becke1993_JCP_5648, Stephens1994_JPC_11623}, and $\omega$B97X~\cite{Chai2008_JCP_84106} functionals, as implemented in {\texttt{LibXC}}~\cite{Lehtola2018_S_1}, are employed for the KS-DFT calculations of the present work.
These used numerical integration with Becke's multicenter scheme~\cite{Becke1988_JCP_2547}, following the approach of Treutler and Ahlrichs~\cite{Treutler1995_JCP_346}.
The default grids (level 3) in {\texttt{PySCF}} were employed, in which the number of radial and angular grid points are (50,302) for H and He, (75,302) for second-row elements, and (80-105,434) for all other atoms.
The grid pruning followed the same procedure as used in {\texttt{NWChem}}~\cite{Apra2020_JCP_184102}.

Additionally, timings for HF calculations using either a first-order implementation in combination with direct inversion in the iterative subspace~\cite{Pulay1980_CPL_393, Pulay1982_JCC_556} (DIIS) acceleration or the second-order trust region optimization of the present work are compared for the homologous series of alkanes with increasingly elongated C--C bond lengths. 
For these calculations, the cc-pVDZ basis set was used for all geometries~\bibnote{These alkane geometries all have C--H bond lengths of $1.09$ \AA\ and angles of $109.5^\circ$.}, and timings were obtained on a single core of a standard Intel Xeon Gold 6230 ($2.10$ GHz) node. 
Furthermore, OpenMP-parallelized timings, obtained using all 40 threads of the above compute node, are reported in Fig. S4 of the SI for the alkanes near their equilibrium geometry in (aug-)cc-pV$X$Z basis sets ($X=\text{D,T,Q}$)~\cite{Dunning1989_JCP_1007, Kendall1992_JCP_6796}. 
All SCF calculations reported herein were started from a superposition of atomic densities (SAD) guess~\cite{Almloef1982_JCC_385, VanLenthe2006_JCC_32}, employing fractionally occupied solutions with the corresponding minimal-energy occupations~\cite{Lehtola2020_PRA_12516}.

For the orbital localizations and symmetrizations of Sects. \ref{sec:orb_loc_results} and \ref{sec:orb_sym_results}, respectively, results are reported for FB and gPM calculations on the buckminsterfullerene (C$_{60}$), circumcoronene (C$_{54}$H$_{18}$), and circumcircumcoronene (C$_{96}$H$_{24}$) systems in both the cc-pVDZ and aug-cc-pVDZ basis sets.
These calculations used the geometries provided in the SI and were started from canonical HF orbitals.
Becke charges, as introduced in Ref. \citenum{Lehtola2014_JCTC_642}, were used for the gPM localization.

\subsection{Self-Consistent Field}\label{sec:scf_results}

Standard first-order methods typically lead to reliable and fast SCF convergence for characteristic organic molecules.
However, these solvers often encounter numerical difficulties in applications to many inorganic compounds, particularly those that contain transition metals. 
While many (but not necessarily all) such systems are insufficiently described by mean-field, single-determinant methods, converged SCF orbitals are still required for generating an initial guess for the active space of multireference approaches. 
The use of a genuine second-order method is thus desirable for these more challenging electronic-structure problems.

\begin{figure}
    \centering
    \includegraphics[width=\linewidth]{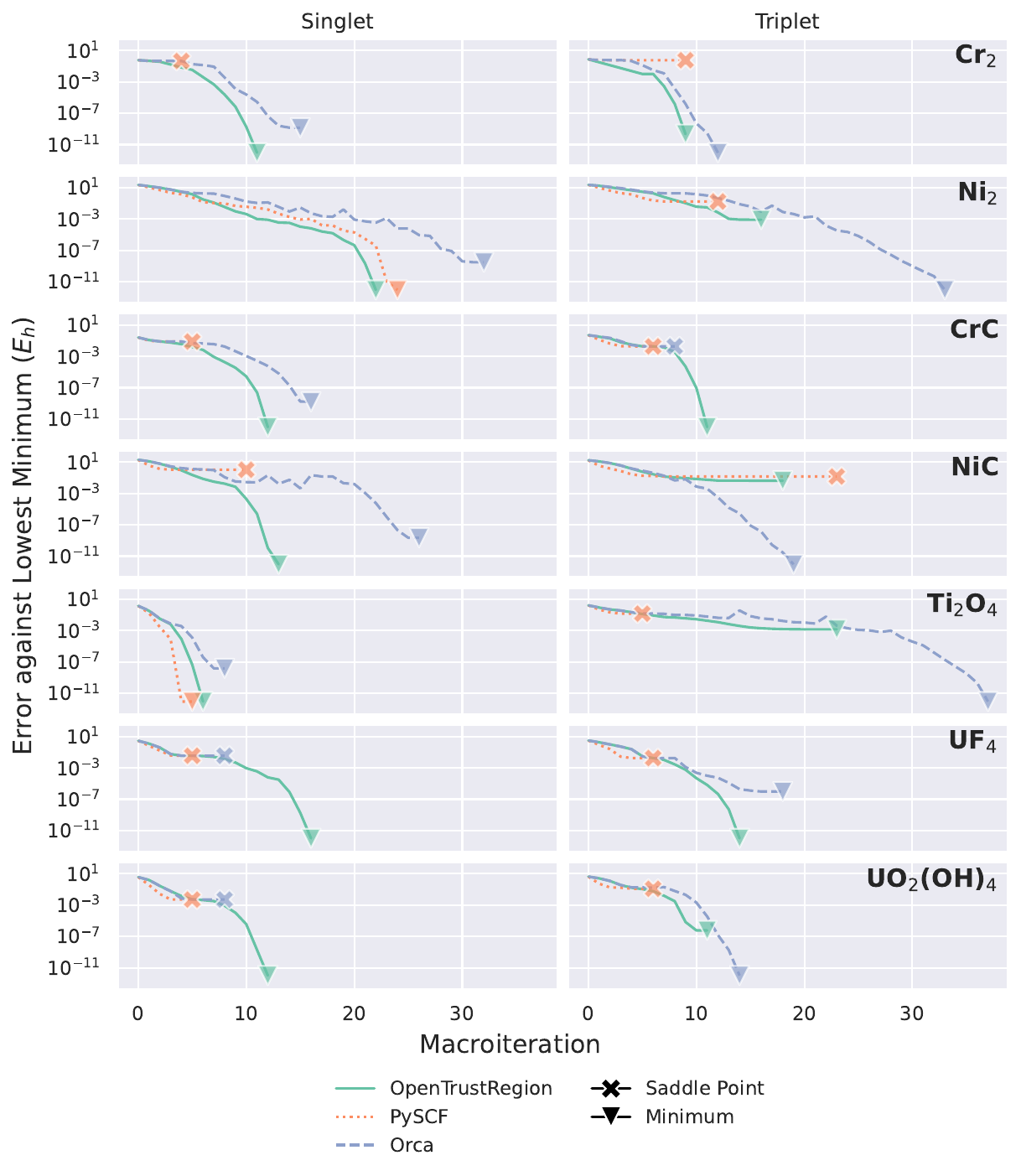}
    \caption{Convergence of second-order orbital optimization of singlet (RHF) and triplet (UHF) states of selected pathological systems. Results are reported for the various implementations in \otr{} (present work), \texttt{PySCF}, and \texttt{Orca}.
    \label{fig:convergence_hf}}
\end{figure}

In the present section, we will compare the performance and robustness of our implementation in \otr{} for restricted and unrestricted SCF to corresponding results obtained using the second-order algorithms implemented in {\texttt{PySCF}}~\cite{sun_pyscf_2017, Sun2020_JCP_24109} (v2.9.0) and {\texttt{Orca}}~\cite{Neese2012_WIRCMS_73, Neese2022_WIRCMS_1606} (v6.0.0).
While the former implementation differs from the algorithm of the present work~\cite{Sun2017__}, the latter closely resembles it~\cite{HelmichParis2021_JCP_164104}.
Convergence profiles for HF calculations on the pathological systems listed in Sect. \ref{sec:comp_details} are presented in Fig. \ref{fig:convergence_hf}.
Additional results comparing the \otr{} and \texttt{PySCF} optimizers for various KS-DFT calculations are presented in the SI.

The group of systems studied in Fig. \ref{fig:convergence_hf} are all known to have multireference character and/or exhibit low-lying excited states, making them susceptible to either failed convergences or optimizations onto saddle points.
As such, these systems serve as a stress test for second-order methods, evaluated first and foremost on their ability to converge onto a (local) minimum.
Since internal stability checks are not available for all methods in every program, we have here assessed convergence by performing the stability analysis in \otr{} for the orbitals obtained from all calculations. 

Overall, the second-order implementation in \texttt{PySCF} (using default parameters) is observed to fail to reliably converge onto a true minimum in most of these calculations, thus contradicting the normal expectation that second-order methods consistently avoid saddle point solutions.
In \texttt{PySCF}, these solutions are recognized as non-Aufbau and appropriate warnings are issued to the user.
As is clear from Fig. \ref{fig:convergence_hf}, such solutions can easily lie less than 1~m\Eh{} above the proper minima identified by other solvers, which goes to emphasize that low-lying, excited-state mean-field solutions pose a real complication in many practical applications. 

The second-order implementation in \texttt{Orca}, on the other hand, generally appears to be much more robust and only a few cases of convergence onto saddle points are observed, e.g., for the singlet states of UF$_4$ and UO$_2$(OH)$_4$ and the triplet state of NiC.
The theoretical similarities of this implementation to ours suggest that these failures are likely caused by the default choice of hyperparameters; for instance, loose thresholds on microiteration convergence may cause the algorithm to miss subtle negative curvature directions, which can potentially result in flawed convergence onto saddle points (cf. the discussion in Sect. \ref{sec:intro}). 

The open-shell UHF calculations in Fig. \ref{fig:convergence_hf} are more prone to multiple local minima, which occasionally leads different solvers to converge onto distinct solutions of different total spin-squared eigenvalues.
This is expected, however, given how second-order algorithms should only guarantee convergence onto a local minimum, not a particular single one.
For the pathological systems studied here, some of the lower-lying solutions are highly spin-contaminated~\cite{HelmichParis2021_JCP_164104}.
Nevertheless, the risk of saddle points in applications of HF and KS-DFT involving first-order solvers to all but the most simple of systems is undeniably real.
We here speculate that such false convergences may even be prevalent across a broad range of published studies and practical applications, as these can only be safely avoided by employing robust second-order optimization methods in combination with stability checks of solutions~\cite{Hughes2025}.

\begin{figure}
    \centering
    \includegraphics[width=\linewidth]{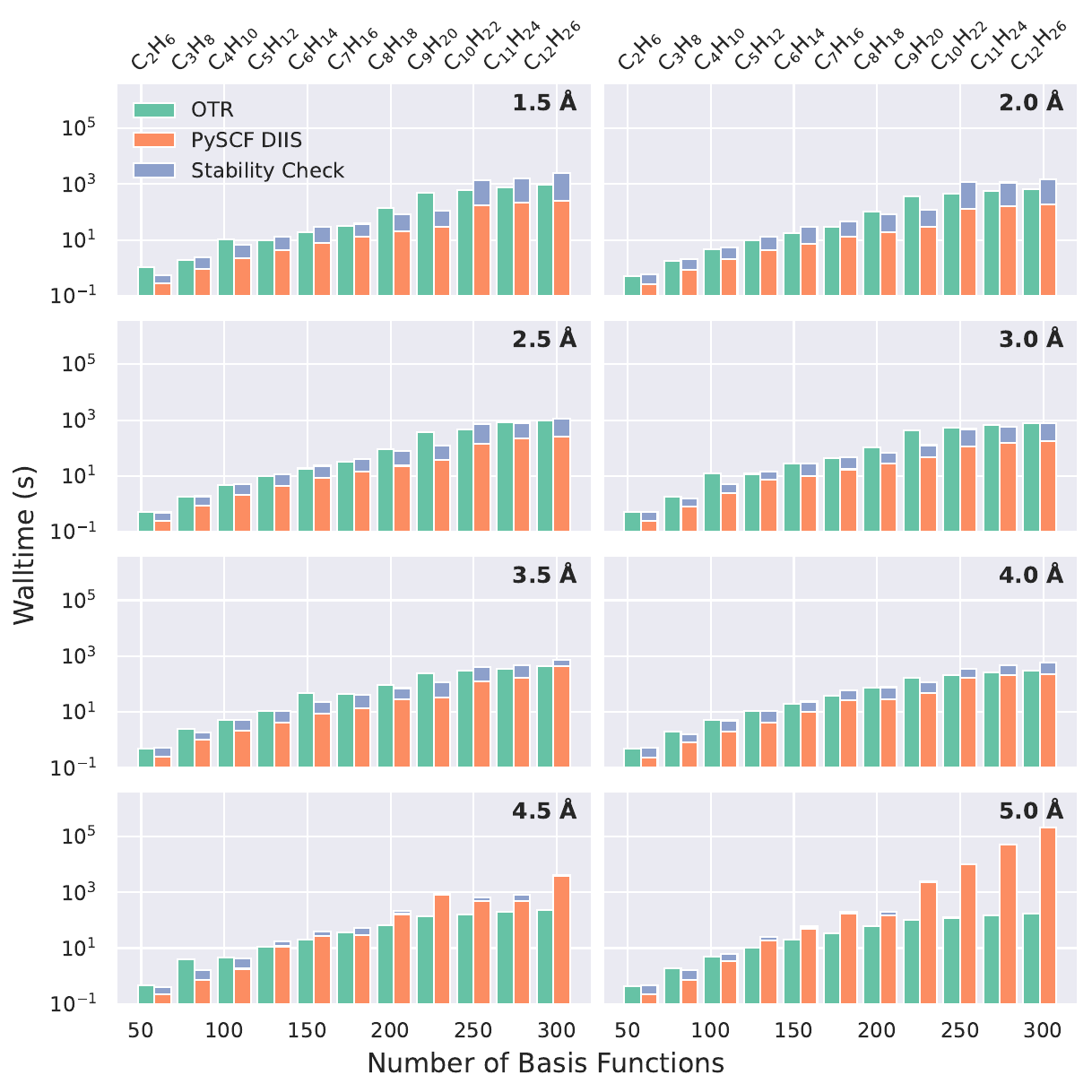}
    \caption{Timings for DIIS (in {\texttt{PySCF}}) followed by an internal stability check and second-order orbital optimization (in \otr{}) for HF calculations on the homologous series of the alkanes with different C--C bond lengths (ranging from $1.5$ to $5.0$ \AA).
    \label{fig:timings}}
\end{figure}

The results in Fig. \ref{fig:convergence_hf} emphasize the need for verifying the stability of a minimum upon convergence.
Our implementation in \otr{}, which yields a local minimum at convergence in all cases, applies stricter defaults---e.g., in the local region---thus enhancing robustness at the expense of more Hessian linear transformations in the SCF optimization.

The need for additional such linear transformations---which are comparable in cost to standard Fock matrix evaluations---is, however, largely negligible when compared to the total cost of (first-order) DIIS followed by an internal stability check, as demonstrated in Fig. \ref{fig:timings}.
For good initial guesses that lie near the quadratic region of the minimum, the slower convergence of the DIIS procedure—combined with the additional cost of Hessian evaluations for stability checks—typically tends to result in higher overall computational cost compared to second-order optimization.
This gap in performance becomes even more pronounced for larger systems and, in particular, more challenging electronic structures, here exemplified by systems with elongated C--C bond lengths, for which DIIS can become more than three orders of magnitude slower than the corresponding second-order procedure.
For small systems, the advantages of the latter are obviously less pronounced, due to the overall low computational cost, but upon an increase in system size, \otr{} is generally competitive with first-order optimization in combination with a subsequent stability check. 
The observation that second-order methods are indeed competitive with first-order methods, whenever these are followed by a subsequent stability check, is corroborated in Fig. S4 of the SI for larger basis sets, possibly augmented by diffuse functions.

The increased cost associated with Hessian linear transformations in the microiterations can be reduced by using the so-called augmented Roothaan--Hall method~\cite{Hoest2008}, and additional progress toward linear scaling can be facilitated by a reformulation of the overall method in terms of AOs, for which local sparsity can be exploited.
Although these natural extensions are outside the scope of the present work, they are subject to ongoing research in our groups.

\subsection{Orbital Localization}\label{sec:orb_loc_results}

The spatial localization of occupied orbitals is generally considered to be a computationally trivial task.
That being said, even for seemingly simple problems, densely clustered extrema with almost identical values of a cost function of choice can result in MOs that fail to match expectation.
In the case of water, for instance, the localized MOs depicted in most textbooks would be comprised of one core orbital and two lone pairs on the oxygen in addition to two symmetric orbitals along the bonds, polarized in the direction of the central oxygen.

When a simple first-order method is used, convergence toward this set of MOs is, however, often not achieved, although starting from an educated atomic guess and employing additional line searches can typically help to avoid convergence onto saddle points.
Nevertheless, reaching this minimum is not guaranteed, and the overall convergence rate of first-order methods can still pose a challenge on complicated optimization landscapes. A more robust optimization method that ensures convergence to a true minimum is therefore preferable.

Robust convergence becomes even more important for the localization of virtual orbitals, as the optimization surface is pronouncedly rugged, and the inclusion of curvature information is required to ensure smooth and reliable convergence.
As an example, the optimization of localized virtual MOs is notoriously difficult for extended, delocalized aromatic systems, particularly in the presence of the diffuse functions of standard augmented basis sets.

\begin{figure}[htb!]
    \centering
    \includegraphics[width=\linewidth]{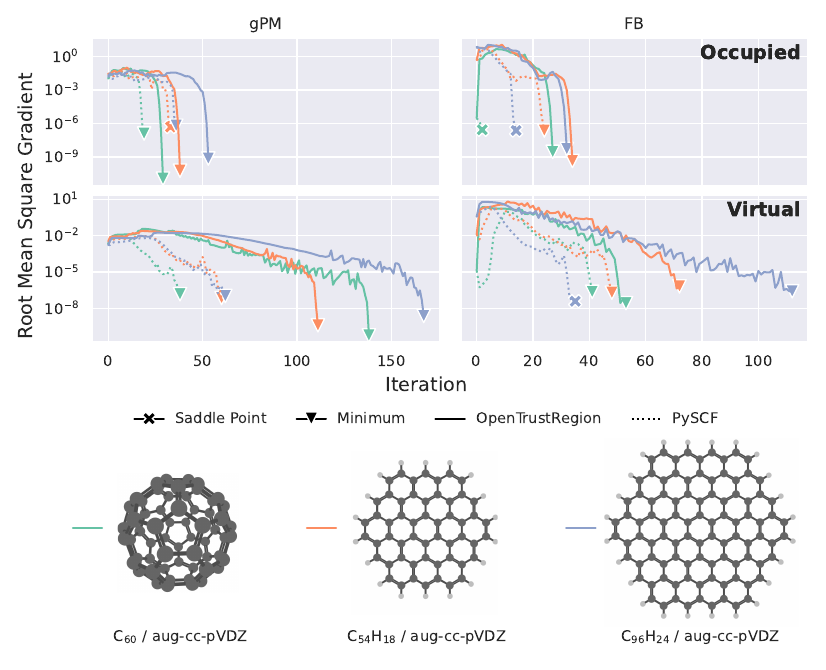}
    \caption{Convergence of gPM and FB MO localization for buckminsterfullerene (green), circumcoronene (orange), and circumcircumcoronene (blue) in an aug-cc-pVDZ basis set.
    \label{fig:convergence_loc}}
\end{figure}
In Fig. \ref{fig:convergence_loc}, we compare the robustness of \otr{} to that of the native {\texttt{PySCF}} solver (using default settings) for the localization of MOs in an aug-cc-pVDZ basis set.
Corresponding results in a cc-pVDZ basis set are provided in the SI.
The occupied orbital localization converges relatively smoothly even for the extended systems considered in Fig. \ref{fig:convergence_loc}.
Nevertheless, the {\texttt{PySCF}} solver occasionally converges onto saddle points.
In the case of C$_{60}$, for instance, the occupied canonical orbitals describe a saddle point on the FB optimization surface, and, starting from these, the solver in {\texttt{PySCF}} will converge onto this solution in a single iteration.
The resulting orbitals are obviously far from spatially local, and convergence onto such a saddle point will be detrimental for the exploitation of locality in subsequent local correlation methods.
Our solver, on the other hand, immediately steps away from the saddle point before eventually converging onto a minimum.
The localization of virtual MOs generally requires more iterations to converge, often in a seemingly erratic manner; nevertheless, the \otr{} solver consistently displays the ability to converge to a true minimum, even when diffuse orbitals are included for these carbon allotropes. 

Not all minima on a localization surface are equal with respect to general locality and symmetry properties.
In principle, symmetric minima should exist that respect the full molecular point group,~\cite{LennardJones1949, LennardJones1950_PRSAMPES_166} but standard localization procedures typically converge onto one of many local minima that only approximately preserve the molecular symmetry.~\cite{eriksen_symlo_jpca_2023}
While a genuine symmetry-constrained optimization method would likely solve this issue, the symmetry properties of the converged orbitals will not be known {\textit{a priori}}.
Recent work by some of us has therefore suggested an alternative procedure to obtain spatially localized orbitals with numerically exact symmetry properties by means of \textit{a posteriori} orbital symmetrization, which is yet another interesting use case of a versatile general orbital optimization library.

\subsection{Orbital Symmetrization}\label{sec:orb_sym_results}

During orbital symmetrization, approximate symmetry invariances and equivalences among a set of spatially localized MOs with respect to symmetry transformations of a molecular point group are enforced by constructing and minimizing a custom cost function~\cite{eriksen_symlo_jpca_2023}.
While the resulting orbitals will no longer correspond to a minimum of the original localization functional, only minor adjustments are typically required, given how the initial localized MOs already exhibit approximate symmetry characteristics.
As a result, the symmetrized orbitals will remain sufficiently local, as gauged by their orbital spreads and tails.

One caveat relates to applications in basis sets with diffuse AOs.
Here, the initial determination of symmetry properties is intrinsically more challenging due to the fact that the localized orbitals may deviate more strongly from symmetry.
Such cases have therefore been excluded from the present work, which merely focuses on illustrating the performance of the \otr{} solver for this optimization task and to demonstrate the versatility of our reusable library for the optimization of rather unconventional objective functions. 

\begin{figure}[htb!]
    \centering
    \includegraphics[width=\textwidth]{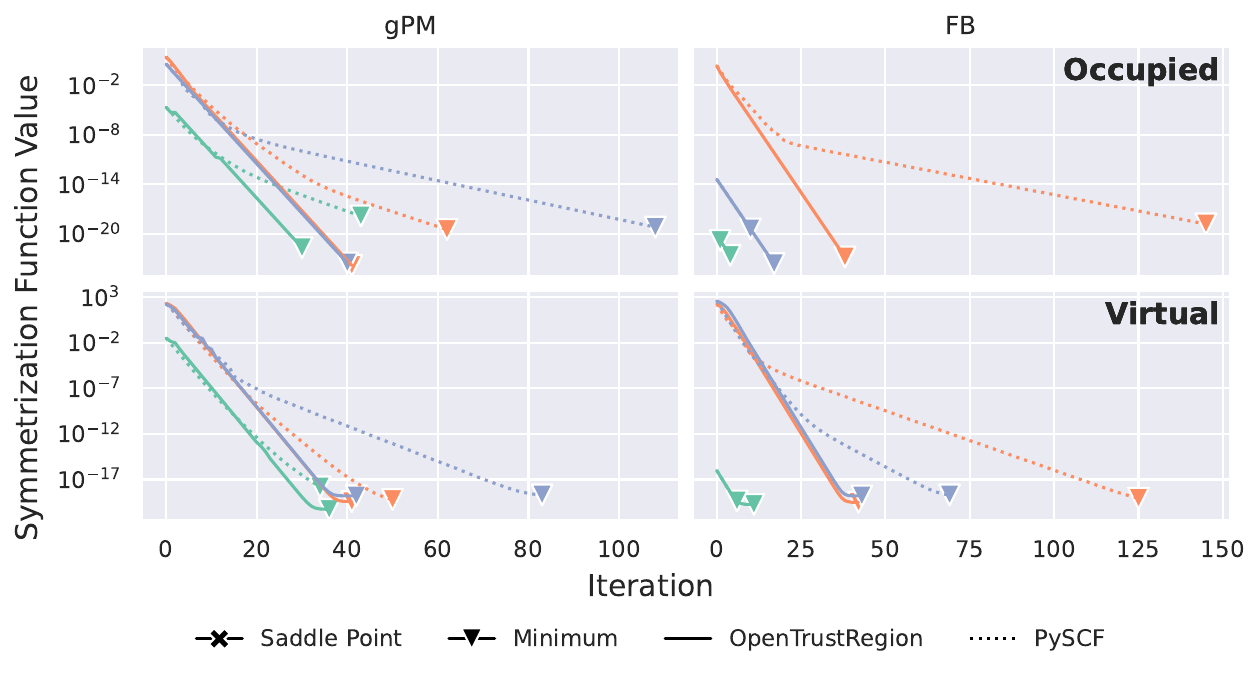};
    \caption{Convergence of the symmetrization of gPM and FB MOs for the same systems as in Fig. \ref{fig:convergence_loc}, albeit in a cc-pVDZ basis set: C$_{60}$ (green), C$_{54}$H$_{18}$ (orange), and C$_{96}$H$_{24}$ (blue).
    \label{fig:convergence_symm}}
\end{figure}
On par with Fig. \ref{fig:convergence_loc}, Fig. \ref{fig:convergence_symm} compares the \otr{} and {\texttt{PySCF}} solvers for the symmetrization of the same localized orbitals, now in a cc-pVDZ basis set (cf. Fig. S4 of the SI).
In comparison to orbital localization, the task of symmetrizing MOs is observed to be considerably more manageable, provided that approximate symmetry properties of the initial orbitals indeed can be correctly identified.
For both sets of localized MOs, saddle point convergence is not observed for any of the three systems in question using either solver, but convergence is generally accelerated using \otr{} over the solver in {\texttt{PySCF}}.

\section{Summary and Conclusions \label{sec:summary}}

We have presented a reusable implementation of the trust region-based second-order orbital optimization algorithm in a new open-source library: \otr{}.
We have demonstrated the versatility of our implementation through applications to both ground-state self-consistent field (HF and KS-DFT), orbital localization, and orbital symmetrization.

Guaranteed and robust convergence to a local minimum in SCF calculations eliminates a number of worries pertinent to many existing implementations, first and foremost being the risk of false convergence onto saddle points.
We demonstrated the difference in speed to conventional methods to be negligible for simple systems, while more pathological systems that are susceptible to convergence issues due to either elongated bonds or transition metal centers will anyway require the robust numerical approach offered by our implementation.

While beyond the scope of the present study, we are currently working on extending the applicability of our library to orbital optimizations for larger molecular systems via the implementation of sophisticated Hessian approximations as well as gradient-based extrapolation methods. Together, these will allow for robust superlinear convergence at only first-order computational cost, while also facilitating the exploitation of sparsity among a set of AOs.

In addition to the applications presented herein to SCF, standard spatial localization (gPM, FB, and ER), and orbital symmetrization, a number of interesting potential further applications of the \otr{} library are worth mentioning.
As a significantly more challenging class of orbital localization, procedures that instead operate by minimizing the fourth-moment orbital spread can be strenuous on account of the fact that the gradient of the cost function often changes by many orders of magnitude during an optimization, thus forcing the trust region to become artificially small~\cite{Hoyvik2012_JCP_224114}.
Likewise, locating stationary solutions for excited states, targeted using orbital-optimized mean-field methods that avoid variational collapses onto the Aufbau configuration~\cite{gilbert_mom_jpca_2008, Barca2014_JCP_111104, Barca2017, Barca2018_JCTC_1501, Levi2020_JCTC_6968, Levi2020_FD_448, Burton2020_JCTC_151, Burton2022_JCTC_1512, Hait2020_JCTC_9, Schmerwitz2023_JCTC_3634}, is known to be numerically demanding.
The same holds for the aforementioned state-specific and state-averaged solutions within MCSCF methods as well as the optimization of spinors in four-component relativistic methods~\cite{Saue2011_C_94}.

Convergence issues may also be encountered for periodic mean-field SCF with a sampling of the first Brillouin zone, particularly so in the study of metals or small band-gap semiconductors, as well as in extensions to the generation of maximally localized and generalized Wannier functions~\cite{Marzari1997_PRB_12847, Jonsson2017_JCTC_460}.
As the optimization functions in these cases depend on orbital densities that are not complex analytic, a generalization of the present library to complex arithmetic ($\mathbb{C}$) is not necessary; the real and imaginary degrees of freedom need to be parametrized separately, and this is ideally handled within interfaces similar to those presented herein.
Finally, emerging methods that involve interacting quantum protons and electrons may lend themselves amenable to our library~\cite{Thomas1969_PR_90, Pavosevic2020_CR_4222, HammesSchiffer2021_JCP_30901}.
The simultaneous optimization of the distinct degrees of freedom has previously been found to lead to improved convergence over two-step algorithms that converge the electronic and protonic problems in alternation~\cite{Liu2022_JPCA_7033}.

Given the ease by which interfaces to various host programs can be written, we expect \otr{} to become widely used in the near future.
Interfaces to \texttt{PySCF}, \texttt{Psi4}, and \texttt{eT} were presented in the course of the present work as examples of codes written and designed in fundamentally different ways, and we bid future community interfaces and extensions to our library most welcome.
As our solver will converge onto the nearest local minimum, it is still important---and perhaps even more so---to employ unambiguous and educated initial guesses in SCF calculations as well as in the context of orbital localization~\cite{Lehtola2020_JCP_144105}.
These will assure that calculations are initiated closer to the locally convex region on a potentially complicated optimization surface and thus both accelerate and ease overall convergence.

\section*{Acknowledgments}

JG and JJE acknowledge financial support from VILLUM FONDEN (a part of THE VELUX FOUNDATIONS) under project no. 37411 as well as the Independent Research Fund Denmark under project no. 10.46540/2064-00007B, both awarded to JJE.
IMH acknowledges funding from the Research Council of Norway through FRINATEK project no. 325574 and support from the Centre for Advanced Study in Oslo, Norway, funding and hosting her Young CAS Fellow research project during the academic years 22/23 and 23/24.
SL thanks the Academy of Finland for financial support under project no. 350282 and 353749.

\section*{Supporting Information}

The supporting information (SI) provides a number of additional results in support of the findings reported in the present study. Table S1 presents a full account of the key hyperparameters used in the calculations behind Figs. \ref{fig:convergence_hf}--\ref{fig:convergence_symm}. Figs. S1--S3 further report results on par with Fig. \ref{fig:convergence_hf}, but obtained using the PBE, B3LYP, and $\omega$B97X functionals instead, respectively. Fig. S4 reports timings on par with Fig. \ref{fig:timings} for HF calculations on the homologous alkane series in extended basis sets, while Fig. S5 reports the convergence of gPM and FB MO localization for the three systems of Fig. \ref{fig:convergence_loc} in a cc-pVDZ basis set. Geometries for these three systems---C$_{60}$, C$_{54}$H$_{18}$, and C$_{96}$H$_{24}$---are further provided as accompanying structure files in {\texttt{.xyz}} format.

\section*{Data Availability}

Data in support of the findings of this study are available within the article and its SI.

\newpage

\providecommand{\latin}[1]{#1}
\makeatletter
\providecommand{\doi}
  {\begingroup\let\do\@makeother\dospecials
  \catcode`\{=1 \catcode`\}=2 \doi@aux}
\providecommand{\doi@aux}[1]{\endgroup\texttt{#1}}
\makeatother
\providecommand*\mcitethebibliography{\thebibliography}
\csname @ifundefined\endcsname{endmcitethebibliography}
  {\let\endmcitethebibliography\endthebibliography}{}


\begin{mcitethebibliography}{180}
\bibitem[Not()]{Note-1}
We note that the diagonal preconditioner in Eq.
  \ref{eq:diagonal_preconditioner} is just one choice among many more elaborate
  options. However, we have found this choice to be generally effective for our
  purposes in all cases where the Hessian matrix is diagonally dominant.
  Furthermore, given that we also rely on the diagonal Hessian for generating
  start vectors (see Sect. \ref{sec:initial-vectors}), the preconditioner is
  readily formed in the algorithm without requiring access to any additional
  data.\relax
\mciteBstWouldAddEndPunctfalse
\mciteSetBstMidEndSepPunct{\mcitedefaultmidpunct}
{}{\mcitedefaultseppunct}\relax
\EndOfBibitem
\bibitem[Not()]{Note-2}
These alkane geometries all have C--H bond lengths of $1.09$ \AA\ and angles of
  $109.5^\circ$.\relax
\mciteBstWouldAddEndPunctfalse
\mciteSetBstMidEndSepPunct{\mcitedefaultmidpunct}
{}{\mcitedefaultseppunct}\relax
\EndOfBibitem
\end{mcitethebibliography}
\end{document}